\documentclass{article}
\usepackage{amsfonts}
\usepackage{amssymb}
\usepackage{graphicx}
\usepackage{amsmath}

\input{tcilatex}

\begin{document}

\title{\textbf{Duality and Fisher zeros in the \\
 2D Potts model on square lattice}}
\author{Marco Astorino$^{1,2}$ \thanks{%
marco.astorino AT gmail.com} , Fabrizio Canfora$^{2}$ \thanks{%
canfora AT cecs.cl} \\
\\
{\small $^{1}$\textit{Instituto de F\'{\i}sica, Pontificia Universidad Cat%
\'{o}lica de Valpara\'{\i}so, Chile}}\\
{\small $^{2}$\textit{Centro de Estudios Cient\'{\i}ficos} (\textit{CECS),
Valdivia, Chile}}}
\maketitle

\begin{abstract}
A phenomenological approach to the ferromagnetic two dimensional Potts model
on square lattice is proposed. Our goal is to present a simple functional form 
that obeys the known properties possessed by the free energy of the q-state Potts model.
The duality symmetry of the 2D Potts model
together with the known results on its critical exponent $\alpha $ allow to
fix consistently the details of the proposed expression for the free energy. The
agreement of the analytic ansatz with numerical data in the $q=3$ case
is very good at high and low temperatures as well as at the critical point.
It is shown that the $q>4$ cases naturally fit into the same scheme and that
one should also expect a good agreement with numerical data. The limiting $%
q=4$ case is shortly discussed.



\end{abstract}

\section*{Introduction}

The Potts model\footnote{%
In the present paper we will only consider spin models on regular square
lattices.} is the most natural generalisation of the two dimensional Ising
model and it is deeply connected with many areas of both statistical physics
and mathematics (for nice detailed reviews see \cite{Wu1}, \cite{Baxter1}, 
\cite{Wu2}, \cite{YG1} and \cite{CARDY}; for the relation with the problem
of colour confinement see \cite{SV82} and the review \cite{PV02}).

It has been very difficult at the moment to compute directly the analytic
free energy of the ferromagnetic Potts model in two dimensions on square
lattice (which is the main object of the present paper) as it was done by
Onsager in the Ising case \cite{O44}. Thus, it will be proposed here a less
direct approach which is, nevertheless, able to give non-trivial analytic
information on the free energy. %
The main purpose of this approach is not to give an exact solution for the free energy, but rather to present a simple functional form that obeys the known properties possessed by the free energy of the q-state Potts model. In particular this method may give a reasonable approximation of the free energy function in all the range of temperatures even the not yet explored ones, i.e. outside the results of high and low temperature or the neighbourhood of critical point.
Following the point of view of the phenomenological Regge theory of scattering (see \cite{Re1} and \cite{Re2};
two detailed reviews are \cite{Co71} and \cite{Ve74}), a reasonable analytic
form for the free energy in terms of one free parameter will be
derived. In the present case it plays the role of Regge's trajectories in high energy physics since it
parametrises some analytical features of the Potts model such as its duality properties and the Fisher
zeros, in an analogue way as the Regge trajectories encode the non-perturbative duality properties of scattering amplitudes, as was first observed in \cite{CPV06}. This can be done by requiring that the sought analytic expression for the
free energy of the 2D Potts model should be compatible (in a suitable sense
explained in the next sections) with the proposal made in \cite{Ca07}, \cite%
{ACMP08}, \cite{ACG09} for the free energy of the three dimensional Ising
model. The proposal of \cite{Ca07}, \cite{ACMP08}, \cite{ACG09} is in a very
good agreement with numerical data so that one should expect that this
approach should provide one with a formula for the free energy of 2D Potts
which fit equally well the available numerical data. It
will be shown that this is indeed the case. A non-trivial by-product of the
present approach is that the proposed free energy of the 2D Potts model has
the locus of the Fisher zeros which coincides with the well known and well
tested conjecture in \cite{FIPOC}.

The paper is organised as follows: in the second section, it will be
discussed a suitable consistency condition (called here "dimensional
compatibility") which allows to derive an analytic ansatz for the free
energy of the 2D Potts model in terms of few $q-$dependent curves. In
the third section, it will be analysed how the known duality symmetry of the
Potts model and its known critical behaviour fix all but one curve. In
the fourth section, the proposed ansatz will be compared with the available
numerical data both at high and low temperatures as well as at the critical
point in the case $q=3$. In the fifth section, the $q\geq 4$ cases will be
discussed. In the sixth section it will be shown how the proposed ansatz
automatically predicts the locus of Fisher zeros consistent with the well
known conjecture.

\section{Dimensional Compatibility}
\label{dimcomp}

It is well known that the main thermodynamical quantities are combinatorial
 in nature.  This fact makes the analysis of three-dimensional 
lattices even more difficult than the two dimensional cases. In this respect, it is a rather surprising
result that one can even imagine to write down a simple explicit functional form for the free energy of the
Ising model in three dimensions which obeys known physical properties (in particular, good agreement with the available numerical data). 
Such simple expression is based on the assumption that
the change in the combinatorial complexity when going from one to two dimensions is formally quite similar to
the analogous change in the combinatorial complexity when going from two to three dimensions. 
One can write the exact Onsager solution for the free energy in two dimensions as a convolution integral of the known
exact free energy in one dimension with a suitable kernel. This integral kernel, which can be found explicitly, is the mathematical
object responsible for the change in the combinatorial complexity when going from one to two dimensions \cite{Ca07}. 
Then, one argues that the change of many combinatorial quantities on hypercubic lattices (such as, for instance, 
the scaling of the number of states of a given energy with the size of the lattice) when going from one to two dimensions
 are similar to the analogous change when going from two to three dimensions (this is somehow confirmed by a comparison with
 the available numerical data \cite{ACG09} , \cite{ACMP08}). Thus, it is not unnatural to assume that a suitable integral kernel exists
such that when one consider its convolution with the Onsager solution one gets the (still unknown) exact free energy of the Ising model
in three dimensions. Our assumption based on the above considerations is that such kernel is as similar as possible to the already known
one allowing to jump from one to two dimensions.\\
The Kallen-Lehmann representation \cite{Ca07} gives rise to an ansatz for the
free energy of the 3DI model of the following form\footnote{%
The parameters appearing in these formulae are related with those appearing
in \cite{Ca07} by the following identities $\zeta _{1}=\nu $, $\zeta
_{3}=1/2=\zeta _{2}$, $\zeta _{0}=\alpha $.} 
\begin{eqnarray}
F_{3D}^{(\zeta _{i},\lambda )}(\beta ) &=&F_{2D}(\beta )+\frac{\lambda }{%
\left( 2\pi \right) ^{2}}\int\limits_{0}^{\pi }dz\int\limits_{0}^{\pi
}dy\cdot  \notag \\
&&\cdot \log \left\{ \frac{1}{2}\left[ 1+\left( 1-\left[ 2\frac{\left(
\Delta (z)-1\right) ^{\zeta _{1}}}{\Delta (z)}\right] ^{\zeta _{2}}\sin
^{2}y\right) ^{\zeta _{3}}\right] \right\} ,  \label{fi7} \\
\ \Delta (z) &=&\left( 1+\left( 1-k_{eff}(\beta )^{2}\sin ^{2}z\right)
^{\zeta _{0}}\right) ^{2},\ \zeta _{0},\zeta _{1},\zeta _{2},\zeta _{3}>0, 
\notag \\
0 &\leq &\left( k_{eff}(\beta )\right) ^{2}\leq 1,\ \ \ 1\leq \Delta (z)\leq
4,  \notag
\end{eqnarray}%
where the values of the parameters in the case of the two-dimensional Ising
model would be $\lambda =1$, $\zeta _{i}=1/2$ for $i=0,..,3$.

Namely, one may find the operator $\mathbf{D}_{1,2}$ (see Eq. (\ref{fode1}))
which "dresses" the trivial one-dimensional solution of the Ising model
giving rise to the Onsager solution. Then, the Kallen-Lehmann free-energy
for the three dimensional Ising model is deduced \cite{Ca07} by modifying $%
\mathbf{D}_{1,2}$ in such a way that the parameters which in $\mathbf{D}%
_{1,2}$ are fixed to be $1$ and $1/2$ (namely, in the two dimensional case, $%
\lambda =1$, $\zeta _{i}=1/2$) become the free parameters. Afterwards, one
can use such a modified operator to dress the Onsager solution obtaining a
useful ansatz for the free energy of the three dimensional Ising model.
To be more precise, it is possible to define a class of operators $\mathbf{D}_{D,q}$ such that when they act on the free
energy $F_{D,q}$ of the $q-$state ferromagnetic Potts model on hypercubic
lattices in $D$ dimensions give rise to the free energy $F_{\left(
D+1\right) ,q}$ of the $q-$state Potts model in $\left( D+1\right) $
dimensions%
\begin{equation}
\mathbf{D}_{D,q}\left[ F_{D,q}\left( \beta \right) \right] =F_{\left(
D+1\right) ,q}\left( \beta \right) .  \label{fode1}
\end{equation}%
To fix the arbitrariness of the above definition it necessary to specify the domain of $\mathbf{D}_{D,q}$ and how it acts.
Since we want that the free energies of reasonable systems belong to the domain of $\mathbf{D}_{D,q}$, the domain of $\mathbf{D}_{D,q}$ will be the class of functions which are smooth on $\mathbb{R}^+$ besides at most a finite number of points in which some of the derivatives of the function may be discontinuous. The variable is the inverse temperature and the singular point (or points) represents the phase transition.
The range of the operator coincides with the domain since, by definition, when this operator acts on the free energy of the Ising model on a hypercubic lattice it generates the free energy of an analogous system in one more dimension. A reasonable way to represent these operators is as integral Kernel \cite{Ca07}: the simplest of these integral kernel can be constructed explicitly by comparing the Onsager solution with the free energy of the one-dimensional Ising model. Then, one can use an integral Kernel of the same functional form to act on the Onsager solution obtaining an approximate ansatz for the functional form of the free energy of the Ising model in three dimensions (the comparison of the ansatz with the numerical data is quite promising \cite{ACMP08} \cite{ACG09}).

\subsection{Consistency condition}

Let $\mathbf{Q}_{D,q}$ be the operators which when applied to the free
energy $F_{D,q}$ of the $q-$state ferromagnetic Potts model in $D$
dimensions give rise to the free energy $F_{D,\left( q+1\right) }$ of the $%
\left( q+1\right) -$state Potts model in $D$ dimensions%
\begin{equation}
\mathbf{Q}_{D,q}\left[ F_{D,q}\left( \beta \right) \right] =F_{D,\left(
q+1\right) }\left( \beta \right) .  \label{fode1.5}
\end{equation}%
The domain of $\mathbf{Q}_{D,q}$ is the class of functions which are smooth on $\mathbb{R}^+$ besides at most a finite number of points in which some of the derivatives of the function may be discontinuous (again the singular points represent the phase transitions of the system).
Also in this case the range of the operator coincides with the domain.
Therefore, being the ranges and the domains of the operators $\mathbf{D}_{D,q}$ and $\mathbf{Q}_{D,q}$ compatible, it makes sense to compose them.
In particular, one can observe that they have to satisfy a sort of commutativity constraint:%
\begin{equation}
\mathbf{Q}_{\left( D+1\right) ,q}\cdot \mathbf{D}_{D,q}=\mathbf{D}_{D,\left(
q+1\right) }\cdot \mathbf{Q}_{D,q}.  \label{almcomm}
\end{equation}

If one supposes that Eq. (\ref{fi7}) is the exact free energy of
the three dimensional Ising model for some values of the parameters,
 the simplest way to satisfy Eq. (\ref{almcomm}) is the following approximate analytic formula for the free energy of the two
dimensional ferromagnetic Potts model: 
\begin{eqnarray}
F_{2D}(q,u) &=&C_{q}+\frac{\lambda (q)}{2\pi }\int\limits_{0}^{\pi }dt\log
\left\{ \frac{1}{2}\left[ 1+\left( 1-\left( k_{2D}(q,u)\right) ^{2}\sin
^{2}t\right) ^{\varsigma \left( q\right) }\right] \right\} ,  \notag \\
\left( k_{2D}(q,u)\right) ^{2} &\leq &1;\ \ \ \ C_{q}=\log \left( 2\frac{%
\exp \left( \beta \right) +\left( q-1\right) }{q}\right) ;\ \ \ \ u=\exp
\left( -\beta \right) ,  \label{klpotts1}
\end{eqnarray}%
where $k_{2D}(q,u)$\ is the function encoding the duality properties of the
model, the function $C_{q}$\ can be found by comparing the high temperatures
expansion of Ising and Potts models in two dimensions (see, for instance, 
\cite{Wu1}), $\lambda \left( q\right) $, $\varsigma \left( q\right) $\ are
two parameters \footnote{$\lambda (q)$ is related to the
overall normalisation of the non-analytic part of the partition function
while $\varsigma \left( q\right) $\ is related to the critical behaviour.}.
It is worth to note that we are still free to add a constant (which depends on q)
to the above free energy. Unfortunately this fact prevents us from using the
Baxter's results on the free energy on the critical point to fix $\lambda
(q) $. It is trivial to verify that the critical point $u_{c}$ corresponding
to the free energy in Eq. (\ref{klpotts1}) is determined by the equation%
\begin{equation}
k_{2D}(q,u_{c})=1  \label{fico1}
\end{equation}%
as one expects on the basis of the Onsager solution.

When $q<4$, the curve $\varsigma \left( q\right) $\ and $k_{2D}(q,u)$\
can be fixed \textit{a priori} using theoretical arguments related to the
duality symmetry and to the known results on the critical behaviour. While we
will fix the normalisation $\lambda (q)$ by a comparison with the numerical data.

As it will be shown in the next sections, $\varsigma (q)$\ is related to the
critical exponent $\alpha \left( q\right) $. Through the well known critical
behaviour of the Potts model, one can get an implicit functional relation
between $\varsigma $ and $\alpha $ 
\begin{equation}
\varsigma \left( q\right) =\Phi \left( \alpha \left( q\right) \right) .
\label{reggetraj1}
\end{equation}%
In the $q>4$ cases (in
which the transition is first order) one can fix \textit{a priori} $%
\varsigma \left( q\right) $ in terms of $\lambda \left( q\right) $ using the
known exact results of Baxter on the latent heat.
Indeed, the above formula may look an ad hoc approximation, or, at least, not very natural at a first glance. 
On the other hand from the arguments at the beginning of section \ref{dimcomp} it stems a constraint on the free energy of the Potts model in two dimensions. Namely, we are interested in finding a functional form that exactly obeys certain properties known to be possessed by the true free energy. So, we search for an expression for the free energy which is compatible with the properties of the integral kernels discussed above.
This gives a further constraint on the form of the free energy of the Potts model in two dimensions. From the combinatorial point of view, this makes the proposed form in Eq. (\ref{klpotts1}) natural. In other words, besides the known constraints on the free energy of the two dimensional Potts model (such as the critical exponent and the duality symmetry), the proposed
form in Eq. (\ref{klpotts1}) is also the simplest compatible with the recursive structure proposed in \cite{Ca07} (which has been proved to be in good agreement with known results \cite{ACG09} , \cite{ACMP08}).
This explains why it is quite useful to look at the recursive structure connecting the Ising models in two and in three-dimensional to obtain an additional bit of information in order to fix the residual arbitrariness left in the choice of the free energy.

\section{Duality and $k_{2D}(q,u)$}

The duality transformation in the case of the two-dimensional Ising model,
was discovered in \cite{KW41} before the exact solution of Onsager \cite{O44}%
. In the case of the Potts model we are considering the following Hamiltonian:
\begin{equation}
                 H=-J \sum_{<i j>} \delta_{\sigma_i \sigma_j} \ \ ,
\end{equation}
 and the duality is
\begin{equation}
\mathit{D}\left( u\right) =\frac{1-u}{1+\left( q-1\right) u}.  \label{duatra}
\end{equation}%
Note that this transformation is not a symmetry of the full free energy \emph{per site} in the thermodynamic limit: it leaves invariant the non-analytic part of the free energy while the trivial term $log(\frac{e^{\beta J} + q -1}{\sqrt{q}})$ is not invariant. Anyway the critical point is determined by the properties of the non-analytic part. The fixed point of the duality transformation $u_{c}=u_{c}\left( q\right) $ is%
\begin{equation}
\mathit{D}\left( u_{c}\right) =u_{c}\Rightarrow u_{c}\left( q\right) =\frac{1%
}{1+\sqrt{q}}.  \label{critpoint}
\end{equation}%
Thus, one has to find a function $k_{2D}(q,u)$ which encodes the duality
properties of the two dimensional Potts model and which reduces to the known
result when $q=2$: that is%
\begin{equation}
k_{2D}(q,\mathit{D}\left( u\right) )=\pm k_{2D}(q,u)  \label{dualcond}
\end{equation}%
where the $\pm $\ signs appear because the Kallen-Lehmann free energy in Eq. (%
\ref{klpotts1}) depends on $\left( k_{2D}(q,\beta )\right) ^{2}$.

The simplest solution (let us call it $\widetilde{k}$) of Eq. (\ref{dualcond}%
) is%
\begin{equation}
\widetilde{k}(q,u)=A\frac{u\left( 1-\left( q-1\right) u^{2}+\left(
q-2\right) u\right) }{\left( 1+\left( q-1\right) u^{2}\right) ^{2}};
\label{corrk2}
\end{equation}%
it can be easily seen that $\widetilde{k}(q,u)$ fulfills Eq. (\ref{dualcond})
for any value of the constant $A$ which we will fix with the normalisation
condition at the critical point in Eq. (\ref{critpoint})%
\begin{eqnarray*}
\widetilde{k}(q,u_{c}) &=&1\Rightarrow \\
\frac{1}{A\left( q\right) } &=&\frac{u_{c}\left( 1-\left( q-1\right)
u_{c}^{2}+\left( q-2\right) u_{c}\right) }{\left( 1+\left( q-1\right)
u_{c}^{2}\right) ^{2}}.
\end{eqnarray*}%
Furthermore the critical point is located at $u=u_{c}$ in Eq. (\ref%
{critpoint}) as it should be and it can be easily seen that when $q=2$ it
reduces to the expression of $k_{2D}$ in terms of the low temperatures
variable $u$.

Indeed, once one has found the simplest $\widetilde{k}$ invariant under
duality transformation and which reduces to the Ising case when $q=2$, one
can construct many more solutions by simply taking functions $f$ of the $%
\widetilde{k}$ in Eq. (\ref{corrk2})%
\begin{equation*}
k_{2D}(q,u)=f\left( A\left( q\right) \frac{u\left( 1-\left( q-1\right)
u^{2}+\left( q-2\right) u\right) }{\left( 1+\left( q-1\right) u^{2}\right)
^{2}}\right)
\end{equation*}%
such that $f\left( x\right) $ has the maximum when $x=1$ and $f\left(
1\right) =1$. The simplest possibility is to consider $f\left( x\right) $ of
the form%
\begin{equation*}
f\left( x\right) =x^{E\left( q\right) }
\end{equation*}

One procedure to determine $E\left( q\right) $ for $q=3$ is to look at the
coefficients of the low temperatures expansions (see \cite{FG}, \cite{GE}).
Using Eq. (\ref{klpotts1}) one can compute the ratios $\delta _{n}$ for very
small $u$ 
\begin{eqnarray*}
\delta _{n}\left( q,u\right) &=&\frac{a_{n}\left( q,u\right) }{a_{n+1}\left(
q,u\right) }, \\
a_{n}\left( q,u\right) &=&\left. \frac{\partial ^{n}F_{2D}(q,u)}{\partial
u^{n}}\right\vert _{u=0}
\end{eqnarray*}%
for some small $n$ and verify that it is possible to fulfil the expected scaling in
the cases $q=3$ with the choice%
\begin{equation}
E\left( 3\right) =2,  \label{exp3y4}
\end{equation}%
so that we will take%
\begin{equation*}
k_{2D}(3,u)=\left( A\left( 3\right) \frac{u\left( 1-2u^{2}+u\right) }{\left(
1+2u^{2}\right) ^{2}}\right) ^{2}
\end{equation*}%
while, of course, the compatibility with the Onsager solution tells that $%
E\left( 2\right) =1$. It is also interesting to observe that consistency
with the $q \rightarrow 1^+$ limit (where the u-dependence disappears) would
suggest $E(1)=0$. \newline
Thus, from now on, we will fix $E\left( 3\right) $ as in Eq. (\ref{exp3y4}).

\subsection{The critical behaviour}

The critical exponent $\alpha \left( q\right) $ for two dimensional Potts
model (when $q\leq 4$) is known to be (see \cite{Pe80})%
\begin{eqnarray}
\alpha \left( q\right) &=&\frac{2\left( 1-2x\right) }{3\left( 1-x\right) }, 
\notag \\
x &=&\frac{2}{\pi }\arccos \left( \frac{\sqrt{q}}{2}\right) ,  \notag
\end{eqnarray}%
where the positive values of $x$ correspond to the tricritical point while
the negative values correspond to the critical point.

One can fix \textit{a priori} the curve $\varsigma \left(
q\right) $ by looking at the critical behaviour of the model: the specific
heat is known to have (see \cite{NS80}, \cite{CNS80}) the following forms
for $q=3$%
\begin{equation}
C_{div}\left( q=3,u\approx u_{c}\right) \varpropto \left\vert u-u_{c}\left(
3\right) \right\vert ^{-1/3}.  \label{dive0.5}
\end{equation}%
On the other hand, the second derivative of the (non-analytic part of the)
free energy $F_{2D}$ in Eq. (\ref{klpotts1}) reads%
\begin{equation*}
\partial _{u}^{2}F_{2D}=-\frac{\varsigma \left( q\right) \lambda \left(
q\right) }{2\pi }\left( \partial _{u}^{2}H\right) \int\limits_{0}^{\pi }%
\frac{\left( \sin ^{2}x\right) \Delta ^{\varsigma \left( q\right) -1}dx}{%
1+\Delta ^{\varsigma \left( q\right) }}+
\end{equation*}%
\begin{equation}
+\frac{\varsigma \left( q\right) \lambda \left( q\right) }{2\pi }\left(
\partial _{u}H\right) ^{2}\int\limits_{0}^{\pi }\frac{\Delta ^{\varsigma
\left( q\right) -2}\left( \sin ^{4}x\right) dx}{1+\Delta ^{\varsigma \left(
q\right) }}\left( \left( \varsigma \left( q\right) -1\right) -\frac{%
\varsigma \left( q\right) \Delta ^{\varsigma \left( q\right) }}{1+\Delta
^{\varsigma \left( q\right) }}\right)  \label{dive1}
\end{equation}%
where%
\begin{eqnarray}
\Delta &=&1-H\sin ^{2}x,\ \ \ \ \ 0\leq \varsigma \left( q\right) <1,  \notag
\\
H &=&\left( k_{2D}\left( u\right) \right) ^{2}.  \notag
\end{eqnarray}%
Near the critical point, $H\approx 1$ and $\partial _{u}H\approx 0$ since,
as it has been already discussed, the critical point is a smooth maximum of $%
k_{2D}\left( u\right) $. For this reason, the most singular term is the
first one:%
\begin{equation}
C_{div}\approx -\frac{\varsigma \left( q\right) \lambda \left( q\right) }{%
2\pi }\left( \partial _{u}^{2}H\right) \int\limits_{0}^{\pi }\frac{\left(
\sin ^{2}x\right) \Delta ^{\varsigma \left( q\right) -1}dx}{1+\Delta
^{\varsigma \left( q\right) }}  \label{dive2}
\end{equation}%
since the divergent term in the integral of the second term in Eq. (\ref%
{dive1}) are compensated by the vanishing first derivative of $H$ at the
critical point (while $\partial _{u}^{2}H$\ is of order $1$). By imposing
that the singular part of the specific heat in Eq. (\ref{dive2}) reproduces
the known critical behaviour in Eq. (\ref{dive0.5}) one gets an implicit
relation between $\alpha $ and $\varsigma $ which fixes $\varsigma \left(
3\right) $ to be 
\begin{equation}
\varsigma \left( 3\right) =0.4  \label{dive3}
\end{equation}%
In the next sections we will draw a picture of the critical part of the free
energy in Eq. (\ref{klpotts1}) against the known results at the critical
point which shows a excellent agreement.

\section{Comparison with numerical data for $q=3$}

An explicit analytic expression for the free energy in Eq. (\ref{klpotts1})
of the two dimensional ferromagnetic Potts model has been constructed in
which theoretical arguments (basically, duality and the known critical
behaviour) can fix everything but one curve $\lambda (q)$. Indeed,
it is easy to see that $\lambda \left( 3\right) $ can fixed in terms of the
numerical expansion data at low temperatures giving rise to a good agreement.
We would like to further emphasise that the main purpose of our work is to obtain an unique approximate functional form for free energy in the widest possible range of temperatures. This is conceptually different and complementary point of view with respect to the known numerical analysis and perturbative expansions. So we have not just assumed the soundness of the low temperature and the critical point numerical studies, but we used them to fix the only free parameter of the ansatz. As it is explained below the precision of our numerical algorithm is not sensitive as the numerical data themselves, neither we pretend to compete with the precision of these methods. Rather we mean to give a global description for the behaviour of the free energy in a unified scheme compatible with the known results on their respective different domains of applicability. Despite the functional simplicity of the free energy in Esq. (\ref{klpotts1}) a low temperature Taylor expansion, which involves many numerical derivations and integrations, using the standard commercial software (available to us) is very inaccurate and anyway beyond the goal of this work.

\subsection{The low and high temperatures}

\begin{figure}[ht]
\label{figlow}
\begin{center}
\includegraphics[angle=0, scale=0.5] {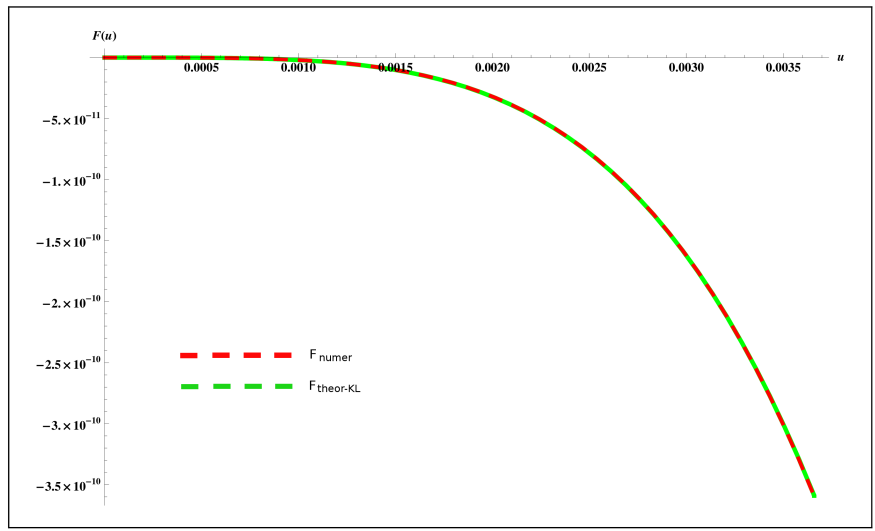} 
\end{center}
\caption{{\protect\small {Numerical $F_{nm}(u)$ and Kallen-Lehmann $%
F_{2D}(u)$ free energies at low temperatures for $q=3$.}}}
\end{figure}

Because of the built-in duality invariance of the model, we will only need
to check the agreement at low temperatures since the agreement at high
follows from duality. Our analysis is based on the references \cite{FG}, 
\cite{GE}: to be more precise, we checked that the normalisation in \cite{GE}
is consistent with (and reduces to) the normalisation of \cite{FG} in the $%
q=2$ case of the two dimensional Ising model. In particular, the expansion
of the partition function in the above references corresponds to only
consider the interesting non-analytic term (neglecting the trivial term $%
\log \cosh 2\beta $ in the Ising case). Therefore, in the Potts case one has
to compare the low temperature expansions for $q=3$ with the second term on
the right hand side of Eq. (\ref{klpotts1}). We consider the low
temperature expansion, up the $14^{th}$ order, valid no more than $%
u_c(3)/100 $, since already in the 2D Ising model the Onsager solution
deviates significantly from the numerical expansion when $u>u_c(2)/100$.  The low temperatures expansion
of the free energy in \cite{FG}, \cite{GE} for the $q=3$ case is 
\begin{eqnarray*}
F_{mc}(u)&=&-\mathrm{Log}(1 + 2 u^4 + 4 u^6 + 4 u^7 + 6 u^8 + 24 u^9 + \\
&& \qquad \ \ + 24 u^{10} + 68 u^{11} + 190 u^{12} + 192 u^{13} + 904 u^{14})
\end{eqnarray*}
and the best value we have found for $\lambda(3)$ is 
\begin{equation*}
\lambda (3) =0.1543
\end{equation*}
For this value of $\lambda(3)$ at low (and therefore high) temperatures the
"precision" of the Kallen-Lehmann ansatz versus the numerical free energy
may be measured in many different ways; for instance one can use the
following: 
\begin{equation*}
p(F_{mc},F_{2D}):=\sqrt{\frac{\int(F_{mc}-F_{2D})^2 du}{ \int (F_{mc})^2 \
du }} \ \ < 1\% \ \ .
\end{equation*}
As one can see from Fig. (\ref{figlow}) the agreement is very good.

\subsection{The critical behaviour}

\begin{figure}[h]
\begin{center}
\includegraphics[angle=0, scale=0.5] {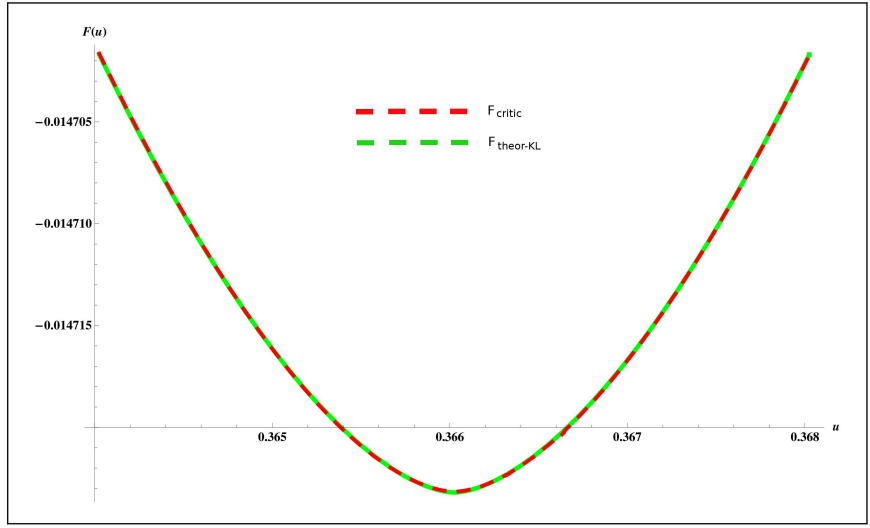}
\end{center}
\caption{{\protect\small {Critical behaviour of free energy $F_{critic}$ and
Kallen-Lehmann $F_{2D}(u)$ near the critical point ($q=3$, $a=-0.0147232$, $%
b=0.677394$)}}}
\label{figcp}
\end{figure}

Because of the simplicity of the free energy in Eq. (\ref{klpotts1}), one
can perform the numerical graph using 
Mathematica$^{{\small \textregistered }}$. The picture \ref{figcp} , in which we
plot the Kallen-Lehmann free energy $F_{2D}(3,u)$ against the critical free
energy 
\begin{equation*}
F_{critic}=a+b\ \big|u-u_{c}(3)\big|^{2-\frac{1}{3}}
\end{equation*}%
in a neighbourhood (sized 1\% of $u_{c}(3)$) of the critical point show a
remarkable agreement. Also in this region the precision $%
p(F_{critic},F_{2D}) $ of the result maintains well below $1\%$.

To get a clearer idea on the precision of the proposed ansatz, one could
compare, using Mathematica$^{{\small \textregistered}}$,
the exact Onsager solution at high temperature versus the corresponding high
temperatures expansion in \cite{FG} in an interval
of $u$ from $0$ to $u_{c}(2)/100$ (the Onsager solution deviates
significantly from the numerical small $u$ expansion when $u>u_{c}(2)/100$).
If one would do this, one would recognise how well the proposed ansatz
describe the available numerical data in the 2D Potts case for $q=3$.

\section{The $q>4$ cases}

The $q>4$ cases are qualitative very different from the cases $q\leq 4$
since, when $q>4$, the phase transition is of first order. Thus, one could
expect that even if the ansatz in Eq. (\ref{klpotts1}) works very well for $%
q<4$, it is not at all obvious if it can also work when $q>4$. Remarkably
enough, it can be shown that the free energy in Eq. (\ref{klpotts1}) does
indeed describe first order phase transition at the critical point $%
u_{c}\left( q\right) $ in Eq. (\ref{critpoint}) provided $\varsigma \left(
q\right) $ becomes negative:%
\begin{equation}
\varsigma \left( q\right) <0.  \notag
\end{equation}

When $\varsigma \left( q\right) $ is negative ($\varsigma \left( q\right)
=-\left\vert \varsigma \left( q\right) \right\vert $), the leading behaviour
of the internal energy corresponding to the free energy in Eq. (\ref%
{klpotts1}) is \ $E_{2D}\left( q,u\approx u_{c}\right) \approx $ 
\begin{equation*}
\approx -\frac{\lambda (q)}{2\pi }\left( 2k_{2D}\partial _{u}k_{2D}\right)
\int\limits_{0}^{\pi }\frac{\left[ \varsigma \left( q\right) \left( 1-\left(
k_{2D}(q,u)\right) ^{2}\sin ^{2}t\right) ^{-\left\vert \varsigma \left(
q\right) \right\vert -1}\sin ^{2}t\right] dt}{1+\left( 1-\left(
k_{2D}(q,u)\right) ^{2}\sin ^{2}t\right) ^{-\left\vert \varsigma \left(
q\right) \right\vert }}\approx
\end{equation*}%
\begin{equation}
\approx -\frac{\lambda (q)}{2\pi }\left( 2k_{2D}\partial _{u}k_{2D}\right)
\int\limits_{0}^{\pi }\left[ \varsigma \left( q\right) \left( 1-\left(
k_{2D}(q,u)\right) ^{2}\sin ^{2}t\right) ^{-1}\sin ^{2}t\right] dt.
\label{klpotts3}
\end{equation}%
Indeed, the derivative of $k_{2D}(q,u)$\ at the critical point is still zero
but the divergence of the integral is stronger so that it gives rise to a
finite contribution, for $u\approx u_{c}$ one has: 
\begin{equation*}
\left\vert \partial _{u}k_{2D}\right\vert \approx 2\varepsilon \ll 1,\ \ \ \
\ k_{2D}\approx 1-\varepsilon ^{2},
\end{equation*}%
while the integral in Eq. (\ref{klpotts3}) is dominated by the region $%
t\approx \pi /2$. Thus, one gets a finite discontinuity of the first
derivative 
\begin{equation*}
E_{2D}\left( q,u\approx u_{c}\right) \approx \mp \frac{\lambda (q)\varsigma
\left( q\right) }{2\pi }\left( 4\varepsilon \right) \int\limits_{-a}^{a}%
\left[ \frac{1}{1-\left( 1-\varepsilon ^{2}\right) \left( 1-\delta
^{2}\right) }\right] d\delta \approx
\end{equation*}%
\begin{equation*}
\approx \mp \frac{2\lambda (q)\varsigma \left( q\right) }{\pi }%
\int\limits_{-\infty }^{\infty }\frac{1}{1+y^{2}}dy=\mp 2\lambda
(q)\varsigma \left( q\right)
\end{equation*}%
where $a$ is a small positive number, the $-$ sign refers to taking the
limit to the critical point from the left and the $+$ from the right.
Therefore the internal energy acquires a finite jump (namely, a non-trivial
latent heat) at the phase transition as it should be. Furthermore, the jump
is symmetric in agreement with the result of Baxter. With the same arguments
one can see that when $\varsigma \left( q\right) $ is positive and less than
one (as it is the case for $q<4$), both left and right derivatives are zero
at the critical point so that the transition is second order. This is a very
interesting fact in itself since it allows to describe with the same
analytical ansatz also the region in which a first order transition takes
place.

The above reasoning tells that to describe the first order region $\varsigma
\left( q\right) <0$\ and to describe the second order region $\varsigma
\left( q\right) >0$. In the second order region it has been shown that $%
\varsigma \left( q\right) $\ is related to the critical exponent $\alpha $.
Below it will be discussed that in the first order region one can fix the
discontinuity of the first derivative of the free energy at the critical
point (which, roughly, is proportional to $\lambda (q)\varsigma \left(
q\right) $) using some exact results obtained by Baxter.

\subsection{Comparison with the Baxter results}

Baxter (see \cite{Baxter1}, for a review see \cite{Wu1}) was able to compute
the latent heat $L\left( q,u_{c}\right) $ at the critical point in the $q>4$
case:%
\begin{eqnarray*}
L\left( q,u_{c}\right) &=&2\left( 1+q^{-1/2}\right) \Delta \left( q\right)
\tanh \left( \frac{\Theta }{2}\right) ,\ \ \ \ \ q>4, \\
\Delta \left( q\right) &=&\prod\limits_{n=1}^{\infty }\left( \tanh \left(
n\Theta \right) \right) ^{2};\ \ \ \cosh \Theta =\frac{\sqrt{q}}{2}.
\end{eqnarray*}%
Thus, one can fix the discontinuity $\Delta E_{c}$ of the first derivative
of the free energy in Eq. (\ref{klpotts1}) at the critical point 
\begin{equation*}
\Delta E_{c}=E_{2D}\left( q,u_{c}^{+}\right) -E_{2D}\left(
q,u_{c}^{-}\right) ,
\end{equation*}%
in terms of the Baxter result:%
\begin{equation}
\Delta E_{c}=L\left( q,u_{c}\right) .  \label{critiq>4}
\end{equation}%
Such equation allows, in principle, to fix one of the two  curves
in terms of the other (for instance, one can choose to express $\lambda
\left( q\right) $ in terms of $\varsigma \left( q\right) $) living us with
only one curve which could be fixed by looking at the numerical expansion
at low temperatures. Unfortunately, at least in the cases $q=5$ and $q=6$
(which we analysed more closely), we have not been able to develop a
suitable software to test at the same time the low temperatures and the
critical behaviour. Even if, at first glance, the numerical problems to be
solved for $q>4$ are similar to the ones which appear for $q<4$, there are
two important differences. The first is that for $q<4$ the known critical
behaviour is only related to $\varsigma \left( q\right) $ which therefore can
be fixed, while for $q>4$ the Baxter result on the latent heat determines $%
\Delta E_{c}$ which is a rather complicated function of \textit{both} $%
\lambda \left( q\right) $ \textit{and} $\varsigma \left( q\right) $ and this
does make the numerical analysis more involved. The second is that, at least
in the cases $q=5$ and $q=6$, the numbers which arise in the low
temperatures expansions are very small and this makes our software extremely
slow.

Nevertheless, we have verified using the software Mathematica$^{{\small %
\textregistered}}$ that looking at the low temperatures only, it is possible to
achieves an almost perfect agreement with the series expansion of \cite{FG}
and \cite{GE}, both for $q=5$ and $q=6$. This is a strong indication that
this framework also works in the case $q>4$ since, because of the built in
invariance under the duality transformation in Eq. (\ref{duatra}), an
excellent matching at low temperatures by construction implies an equivalent
agreement at high temperatures. Thus, taking into account that in the $%
\varsigma \left( q\right) <0$ region the free energy in Eq. (\ref{klpotts1})
has a first order phase transition, one should expect that the
present method provides an explicit analytic description of the
free energy both for $q<4$ and for $q>4$\ in terms of only two $q-$dependent curves\footnote{%
Furthermore, at least one of these two curves can be fixed \textit{a
priori} analytically using known analytical results at the critical point
(see Eqs. (\ref{dive3}) and (\ref{critiq>4})).}, in excellent agreement with
numerical expansion data.

\subsection{The $q=4$ case}

The $q=4$ case is the more delicate. The first obvious reason is that $q=4$
is the boundary between the range in which the model exhibit a second order
phase transition ($q\leq 4$) and the range in which the transition is first
order (that is, $q>4$). As a matter of fact, for $q=4$: the specific heat
singularity has logarithmic correction (see \cite{NS80} or \cite{CNS80}):%
\begin{equation}
C_{div}\left( q=4,u-u_{c}\right) \approx \frac{\left\vert
u-u_{c}(4)\right\vert ^{-2/3}}{\log \left\vert u-u_{c}(4)\right\vert }.
\label{logdiv}
\end{equation}%
The present formalism provides one with a very natural mechanism for the
arising of logarithmic corrections.

Assuming that $\varsigma \left( q\right) $ is a continuous function of 
$q$, then one could argue that $\varsigma \left( 4\right) =0$ since in the
first order region $\varsigma \left( q\right) <0$\ while in the second order
region $0<\varsigma \left( q\right) <1$. The interpretation of the
last sentence is that, when $\varsigma \left( q\right) =0$, the present
framework provides one with the appearance of the expected "nested"
logarithm inside the free energy: 
\begin{equation}
F_{2D}(4,u)=C_{4}+\frac{\lambda (4)}{2\pi }\int\limits_{0}^{\pi }dt\log
\left\{ \frac{1}{2}\left[ 1+\Xi \left( q,u\right) \log \left( 1-\left(
k_{2D}(4,u)\right) ^{2}\sin ^{2}t\right) \right] \right\} .  \notag
\end{equation}%
Unfortunately, we have still not found a theoretical argument to fix \textit{%
a priori} $\Xi $ in an analytic way.\ Nevertheless, it is worth to stress
that the above formula does give rise \textit{automatically} to a second
order phase transition with a logarithmic correction of the type in Eq. (\ref%
{logdiv}) and, therefore, to find theoretical arguments able to fix $\Xi
\left( q,u\right) $\ is an interesting open problem.

\section{Fisher zeros}

A powerful theoretical tool is the analysis of the Fisher zeros \cite{Fi68}
in which the inverse temperature $\beta $ is extended to the whole complex $%
\beta $-plane (in the same way as Yang and Lee complexified the magnetic
field \cite{LY}). By looking at the distribution of zeros of the partition
function in the complex $\beta $ plane, one can determine the universal
amplitude ratio $A_{+}/A_{-}$ of the specific heat and write down simple
expressions (which only involve the density of zeros and the angle which the
line of zeros form with the real $\beta $ axis at the critical point) for
the free energy and the specific heat close to the critical point; see \cite%
{Suzuki} \cite{Abe}. Such tools are also useful when analysing the strength
of the phase transitions (see for instance \cite{Kenna},\ \cite{Kenna2}, 
\cite{Ken98}\ and references therein). Therefore, the Fisher zeros contain
very deep non-perturbative informations on the corresponding systems.

It is expected that also in the case of the two dimensional Potts model the
Fisher zeros should lie on a circle: strong theoretical as well as numerical
evidences have been provided in \cite{FIPOC} (for some more recent evidences see \cite{salas1}, \cite{salas2} and references therein). The circle is given by $|x| = 1$, where $x = v/\sqrt{q}$ and $v = e^{\beta} - 1 = u^{-1} - 1$.
To study the locus of the Fisher zeros, it is convenient to express $\widetilde{k}(q,u)$
as follows: 
\begin{equation}
\widetilde{k}(q,u)=4\frac{uD\left( u\right) }{\left( u+D\left( u\right)
\right) ^{2}} \ .  \label{duexp}
\end{equation}%
It is easy to observe from (\ref{duexp}) that\footnote{%
One may notice that in the above expression in Eq. (\ref{duexp}) for $%
\widetilde{k}(q,u)$ the dependence on $q$ is implicit in the duality
transformation $D(u)$. Furthermore, $\widetilde{k}(q,u)$ reduces to the
known expression for $q=2$. Therefore, when expressed in terms of $u$ and $%
D(u)$, the equation determining the locus of Fisher zeros is formally
exactly the same as in the Ising case provided one replaces the duality
transformation of the Ising case with the corresponding Potts duality
transformation.} $\widetilde{k}(q,u)  \leq  1$ and $\widetilde{k}(q,u) = 1$ iff $u=D(u)$; this condition identifies the real fixed point of the duality map.
On the other hand according to the conjecture, the locus of Fisher zeros of the 2D Potts model on square lattice is the analytic extension of the equation $u_{c}=D(u_{c})$. Therefore our proposal is consistent with the conjecture providing further support to this framework.

\section{Conclusions and perspectives}

A phenomenological approach to the ferromagnetic two dimensional Potts model
on square lattice has been developed. After introducing the $D-$dressing and
the $q-$dressing operators $\mathbf{D}_{D,q}$\ and $\mathbf{Q}_{D,q}$, it
has been described how the compatibility between $\mathbf{D}_{D,q}$\ and $%
\mathbf{Q}_{D,q}$ allows one to write down an explicit analytic ansatz for
the free energy in terms of one free parameter (for each q). The duality
symmetries of the 2D Potts model together with the known theoretical results
on its critical exponent allow to fix \textit{a priori} all but one
curve. The agreement of the proposed analytic free energy with low and high temperatures expansion as
well as the critical point is excellent for $q=3$. For $q=5$ and $q=6$ one can also see that the
agreement with numerical data at low and high temperatures is also very
good but we have not been able to test the corresponding critical points
because of some subtle numerical problems. Nevertheless it has been proved
that the corresponding phase transition when $\varsigma <0$ is first order.
The $q=4$ case remains basically opens but we have some indications that the
present framework is also able to capture important features of such subtle
case since it predicts automatically logarithmic correction to the power law
divergence of the specific heat.\newline

\section*{Acknowledgements}

{\small {The authors want to give a very warm thank for his illuminating
comments and suggestions to G. Giribet which participated to the early stage
of this project. The authors want also to thank P. Butera for important
bibliographic remarks. This work was supported by Fondecyt grant 11080056.
The Centro de Estudios Cient\'{\i}ficos (CECS) is funded by the Chilean
Government through the Millennium Science Initiative and the Centers of
Excellence Base Financing Program of Conicyt. CECS is also supported by a
group of private companies which at present includes Antofagasta Minerals,
Arauco, Empresas CMPC, Indura, Naviera Ultragas and Telef\'{o}nica del Sur.}}

{\small \bigskip }

\end{document}